\RequirePackage{tikz,pgfplots}
\documentclass[pdflatex,sn-mathphys]{sn-jnl}% Math and Physical Sciences Reference Style
\usepackage[T1]{fontenc}
\usepackage{amsmath}
\pgfplotsset{compat=newest}
\pgfkeys{/pgf/number format/.cd,1000 sep={}}
\usetikzlibrary{decorations.markings}
\usetikzlibrary{plotmarks}
\usepgfplotslibrary{groupplots}

\newcommand{\ut}[1]{\,\mathrm{#1}}
\newlength\figureheight 
\newlength\figurewidth 
\jyear{2021}%

\theoremstyle{thmstyleone}%
\theoremstyle{thmstyletwo}%

\theoremstyle{thmstylethree}%

\begin{document}

\title[Article Title]{Magnetization in superconducting corrector magnets and impact on luminosity-calibration scans in the Large Hadron Collider}


\author*[1]{\fnm{Agnieszka} \sur{Chmieli\'{n}ska}}\email{agnieszka.chmielinska@cern.ch}
\author[1]{\fnm{Lucio} \sur{Fiscarelli}}
\author[2]{\fnm{Michi} \sur{Hostettler}}
\author[3]{\fnm{Witold} \sur{Kozanecki}}
\author[1]{\fnm{Stephan} \sur{Russenschuck}}
\author[1]{\fnm{Ezio} \sur{Todesco}}

\affil[1]{\orgdiv{Technology Department}, \orgname{CERN}, \orgaddress{\street{Esplanade des Particules 1}, \postcode{1211} \city{Meyrin}, \country{Switzerland}}}

\affil[2]{\orgdiv{Beams Department}, \orgname{CERN}, \orgaddress{\street{Esplanade des Particules 1}, \postcode{1211} \city{Meyrin}, \country{Switzerland}}}

\affil[3]{\orgdiv{IRFU-DPhP}, \orgname{CEA/Université Paris-Saclay}, \postcode{91191} \city{Gif-sur-Yvette}, \country{France}}


\abstract{Superconducting accelerator magnets have a nonlinear dependence of field on current due to the magnetization associated with the iron or with persistent currents in the superconducting filaments. This also gives rise to hysteresis phenomena that create a dependence of the field on the powering history. Magnetization effects are of particular importance for luminosity-calibration scans in the Large Hadron Collider, during which a small number of Nb--Ti superconducting orbit correctors are excited at low field and with frequent flipping of the sign of the current ramp. This paper focuses on the analysis of special measurements carried out to estimate these nonlinear effects under the special cycling conditions used in these luminosity scans. For standard powering cycles, we evaluate the effect of the main magnetization loop; for complex operational schemes, magnetization-branch transitions occur that depend on the details of the current cycle. The modelling of these effects is not included in the magnetic-field prediction software currently implemented in the LHC control system; here we present an approach to predict the transitions between the main magnetization branches. The final aim is to estimate the impact of magnetic hysteresis on the accuracy of luminosity-calibration scans.  

}

%%================================%%
%% Sample for structured abstract %%
%%================================%%




\maketitle

\section{Introduction}\label{sec1}

Superconducting accelerator magnets exhibit nonlinear effects due to magnetization, either of the iron or of the superconducting filaments~\cite{Green:1971, Wilson:1987, Schmuser:1991, Russenschuck}, and cause magnetic hysteresis that affects the magnetic transfer function at low excitation levels. They also induce a dependence on the powering history that is quite complex to be modelled. Even though these effects for the main magnets are on the order of 0.1\% at injection current, they have to be carefully taken into account in the field model for operating the CERN Large Hadron Collider (LHC)~\cite{Sammut:2006}. The dependence on the powering history is mostly eliminated by imposing a pre-cycling strategy to all magnets. On the other hand, these effects are usually neglected for the corrector magnets, since they are a second order effect on the corrective action. For this reason, the LHC field model neglects magnetization and hysteresis phenomena in corrector magnets.

Luminosity-calibration scans~\cite{Grafstrom:2015} at the LHC require a careful consideration of the nonlinear effects related to magnetization in corrector magnets. During such scans, the superconducting orbit correctors MCBC and MCBY~\cite{Bruning:2004} are used to sweep the colliding beams transversely across each other at the interaction point (IP) in the center of the particle-physics detectors. During these luminosity-calibration scans, the magnets operate at low current levels and are subject to local field variations due to frequent inversions of the current-ramp direction. Recently, the analysis of the errors associated with the absolute calibration of the luminometers has identified the magnetization of MCBC and MCBY and the associated effects on the beam displacement as a potential source limiting the precision of the scans~\cite{Atlas, CMS}. This motivated the need for dedicated magnetic measurements of these correctors aiming at a precision in the relation current/field on the order of 0.1$\%$.

This paper reports the special measurements on the corrector magnets done to reproduce their behaviour during the special cycling conditions that are used in the luminosity-calibration scans. We then propose a methodology on how to fit the measured magnetic hysteresis so as to achieve the required accuracy on the magnetic field prediction for a given excitation history. The measured magnetization effects have been parameterized using a subset of the measurements, their impact on actual beam displacements during luminosity scans has been modelled using that parameterization, and the results have been compared with beam-displacement measurements~\cite{Atlas} at the ATLAS IP.

This paper is organized as follows. The physics background is outlined in Sect.~\ref{sec2}, including an overview of magnetization effects in superconducting filaments, the description of the LHC magnetic-field prediction system and the relevant aspects of the luminosity-calibration protocol. Section~\ref{sec3} describes the set of measurements used to  investigate the magnetic hysteresis. The results of the magnetic measurement campaign are discussed in Sect.~\ref{sec4}, and a novel numerical model describing the branching of magnetic hysteresis is developed in Sect.~\ref{sec5}. In Sect.~\ref{sec6}, the results of beam-displacement simulations based on the aforementioned model are compared with ATLAS measurements. The conclusions are drawn in Sect.~\ref{sec7}.

%% \section{\label{sec2}Background}
\section{\label{sec2}Physics background}

\subsection{\label{sec2:1}Magnetization from persistent currents}

Even though the magnetization in superconducting magnets has two main sources, namely the iron magnetization and the superconducting filament magnetization, in the corrector magnets MCBC and MCBY, the second one dominates over the first one. In this paragraph we give the theoretical basics for modelling superconducting filament magnetization.

The phenomenological description
of hard superconductors is based on an electrical
conductor with an $E(J)$ characteristics given by
the power law~\cite{Russenschuck}:
\begin{equation}\label{ej0}
\frac{E}{E_\mathrm{c}} = \left(\frac{J}{J_\mathrm{c}} \right)^n, 
\end{equation}

\noindent where $E := \lvert \textbf{E} \rvert$, $J := \lvert \textbf{J} \rvert$ are the absolute values of the 
electric field and current density, $E_\mathrm{c}$ and $J_\mathrm{c}$ denote the threshold electrical field criterion and the critical current density respectively, and $n$ is the resistive-transition index. From
Eq.~(\ref{ej0}) we obtain in vector form:

\begin{equation}\label{ej}
\textbf{E} = E_\mathrm{c} \left( \frac{J}{J_\mathrm{c}} \right)^{n-1} \frac{ \textbf{J} }{J_\mathrm{c}}.
\end{equation}
For current densities close to $J_\mathrm{c}$, flux creep gives rise to an electric field that varies exponentially with $n$. The resistive-transition index is as large as 50 for multifilamentary Nb--Ti wires.
Measurements for LHC strands give field-dependent values of
$n$ = 42 at 10~T and $n$ = 48 at 8~T.

By definition, the critical current density
is reached when the electric field attains 
1~\textmu $\mathrm{V\,cm^{-1}}$. It can therefore be taken as the constitutive equation for hard superconductors~\cite{Ekin2006}. It also serves for the modeling of field penetration into the specimen by the process of nonlinear diffusion.

Since the resistivity of hard superconductors is nearly a
step function, it has been postulated that the current density in a hard superconductor is always either zero or equal to the critical current density.  This rule is known as  the \textit{critical state model} (CSM)~\cite{Bean:1962}. A time-transient magnetic field induces an electric field at the surface of the conductor, which gives rise to a current density slightly above $J_\mathrm{c}$, so that the resistive
voltage matches the electric field. When the field sweep stops, the current in the slab decays until $J_\mathrm{c}$ is reached.

For a magnetic field applied at the strand surface, a simple way to calculate the superconductor magnetization is to apply Amp\`{e}re's law, which yields a linear profile of the magnetic field within the strand. From the shielding-current distribution in the strand, the magnetic moment and its effect on the magnetic field in a superconducting magnet can be  straightforwardly  calculated~\cite{aleksa2002,aleksa2004,Vollinger:2002}.

\subsection{\label{sec2:2}The magnetic field model in the LHC}

In particle accelerators, an accurate knowledge of the magnetic field generated by the superconducting magnets is required for transverse and longitudinal beam control~\cite{bryant1993principles}. 
At the LHC, the \textit{Field Description for the LHC} (FiDeL) is used 
for determining for each class of magnets the current level for the required field strength. The model is based on the identification and decomposition of static and dynamic components that contribute to the total field in the magnet aperture. FiDeL is based on fitting a series of magnetic measurements with functions that keep the physics of the different components~\cite{Sammut:2006, Sammut:2007, Sammut:2009}.

The FiDeL model uses different levels of complexity, starting from a linear dependence on the current and adding terms to describe the nonlinear effects, such as magnetization at low fields or saturation of the iron yokes at high fields. 
The main challenge in the operation of the magnets is to find a model of the magnetic transfer function able to predict the nonlinear effects with an acceptable error. 
The complexity of the model therefore depends on the type of magnet and its optical function. 

The LHC field model neglects magnetization effects in all corrector magnets, because these lie well below the accuracy level required by standard operation~\cite{FiDeL_MCBC:2009}, \textit{i.e.} below 1$\%$ relative to the main field. However, the special use case of MCBC and MCBY for the luminosity-calibration scans described in Sec.~\ref{sec2:3}, leads to much stricter requirements. In order to remain negligible with respect to the systematic-uncertainty budget of these scans, which at HL-LHC is as tight as 0.6\% from all sources combined, the actual field variation must remain linear with respect to the dialed-in field change to within 0.1\%  of the maximum field excursion during the scans; it must in addition remain reproducible, at the same level of accuracy, from one scan to another. Achieving this level of precision lies beyond the capability of the present FiDeL model of LHC corrector magnets.  Moreover, FiDeL predicts only standardized cycles, \textit{i.e.} ramping from injection to maximum current, and then ramping down to injection. During luminosity scans however, the powering schemes of the MCBC and MCBY magnets are more complex, with frequent reversal of the current-ramp at low current levels. This is why we will carry out a special investigation in this domain, extending the initial scope of FiDeL.

\subsection{\label{sec2:3}Luminosity-calibration scans}
At the LHC, the calibration of the experimental luminometers, {\em i.e.} the determination of their absolute luminosity scale, is based on dedicated {\em van der Meer} (vdM) scans, whereby the absolute luminosity is inferred, at one point in time, from the measurable parameters of each colliding-bunch pair~\cite{Grafstrom:2015}. By comparing the known luminosity delivered at the peak of the scan, where the colliding bunches are perfectly centered on each other in the transverse plane, to the corresponding counting rate reported by the luminometer, the proportionality constant between these two quantities can be determined to sub-percent accuracy~\cite{Atlas}. To minimize systematic uncertainties on the luminosity scale, these scans are typically performed under carefully controlled conditions and with beam parameters optimized for the purpose.

\par
For a given colliding-bunch pair $b$, combining the measured bunch populations $n_1$ and $n_2$ (which do not concern us here), with the horizontal and vertical convolved beam sizes $\Sigma_x$ and $\Sigma_y$ determined by the vdM method, yields the absolute luminosity ${\mathcal L}_b$ associated with that bunch pair:
\begin{equation}
{\mathcal L}_b = \frac{ f_{\mathrm r} n_1 n_2 }{{2\pi \Sigma _x \Sigma _y
}},
\end{equation}
where $f_{\mathrm r}$ is the machine revolution frequency. 

\par
\begin{figure}[htb!]
\centering
{\includegraphics[scale=0.57]{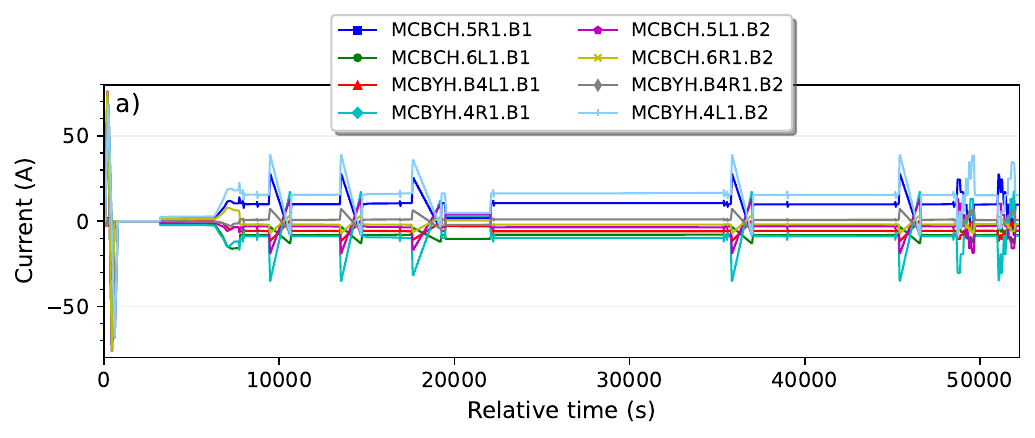}} \\
{\hspace*{0.20cm} 
\includegraphics[scale=0.57]{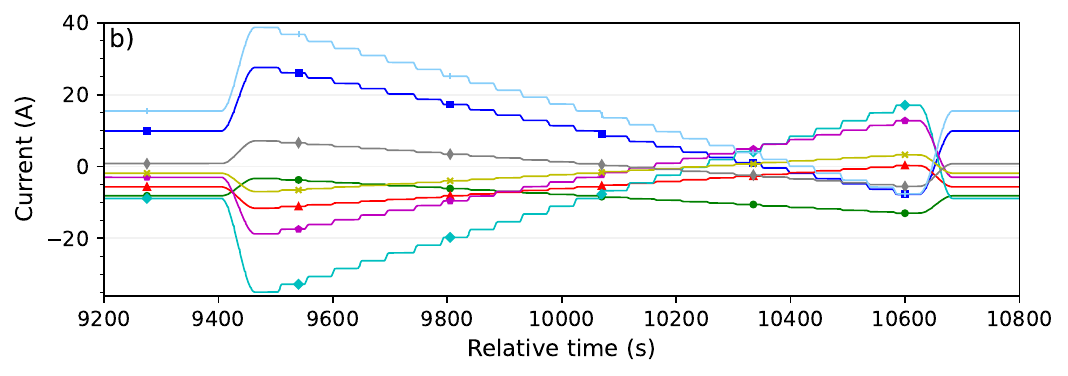}}%
 \quad 
\includegraphics[scale=0.55]{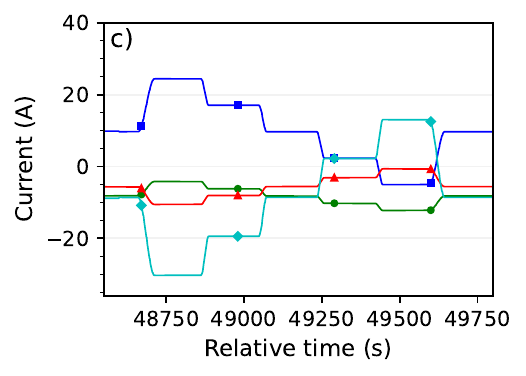}
\includegraphics[scale=0.55]{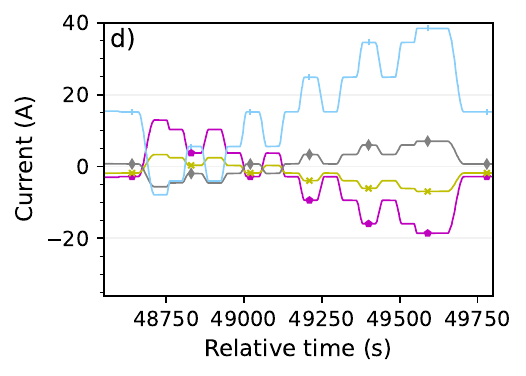}
 \caption{\label{fig:6016}%
Powering sequence of MCBC and MCBY magnet circuits executed during luminosity-calibration scans in order to provide horizontal deflection (LHC fill 6016): (a) a pre-cycle up to nominal current followed by a sequence of different scans, (b) a single vdM scan, (c) excitation history of the correctors controlling the target beam during an LSC scan, (d) excitation history of the correctors associated with the witness beam during an LSC scan. }%
\end{figure}

The convolved beam sizes $\Sigma_x$ and $\Sigma_y$ are determined~\cite{Grafstrom:2015} by scanning the beams transversely across each other in opposite directions, both in the horizontal and in the vertical plane, while simultaneously measuring the collision rate as a function of the transverse beam separation: these are known as vdM or {\em beam-separation} scans. These scans typically cover 25 steps that span a range of $\pm 6\, \sigma_b$ in beam separation, where $\sigma_b$ is the nominal transverse single-beam size at the collision point. This amounts to $\pm 3\, \sigma_b$, typically several hundred micrometers, for each beam and each plane separately. The corresponding excursions in orbit-corrector current are illustrated by the  five leftmost scans shown in Fig.~\ref{fig:6016}a; an expanded view of the first of these scans is shown in Fig.~\ref{fig:6016}b. 

\par
The ($\Sigma_x$, $\Sigma_y$) measurements above require the knowledge of the absolute length scale, {\em i.e.} of the actual beam displacement that corresponds to a given nominal beam displacement dialed into the LHC control system. This beam displacement is determined by a {\em length-scale calibration} (LSC) scan. The principle is to move the beams parallel to each other, keeping them in perfect head-on collision, and to calibrate each of the four closed-orbit bumps (two beams, two planes)  against the absolute displacement of the luminous centroid measured by the ATLAS (or ALICE, CMS or LHCb) tracking system using reconstructed collision vertices.

\par
During the LSC scan, the {\em target} beam, {\em i.e.} that affected by the closed-orbit bump under calibration, is scanned over at least five equally spaced positions that span $\pm 3\, \sigma_b$ in nominal single-beam displacement; the scanning range and direction (positive to negative, or vice-versa) are required to be identical to those used during the beam-separation scans. The requirement for the two beams to remain in head-on collision is satisfied by performing, at each step, a three-point, beam-separation miniscan of the {\em witness} beam around the nominal position of the target beam, fitting the resulting curve of luminosity vs. luminous-centroid position, and interpolating the luminous-centroid position to that of maximum luminosity and beam overlap~\cite{Atlas}. This procedure is illustrated in the two rightmost scans of Fig.~\ref{fig:6016}a. The expanded view of Fig.~\ref{fig:6016}c illustrates the excitation history of the correctors associated with the target beam (in this case beam 1), such as MCBCH.5R1.B1. That of the correctors associated with the witness beam, for instance MCBYH.4L1.B2, is noticeably different (Fig.~\ref{fig:6016}d).

\section{\label{sec3}Measurement strategy}

The goal of the magnetic measurements is to quantify and understand the nonlinear effects from magnetization under the powering conditions that occur during luminosity-calibration scans. To assess the impact of cycles other than those performed during standard operation, we conducted a dedicated measurement campaign. Due to the wide range and the complexity of machine cycles, the selection of the suitable powering schemes was a major challenge. Hence, we programmed a large set of measurements to reproduce the relevant features of the magnet cycles used in vdM scans. 

\subsection{\label{sec3:1}Measurement setup}
\begin{figure}[htb!]
\centering
{\includegraphics[width=0.75\linewidth]{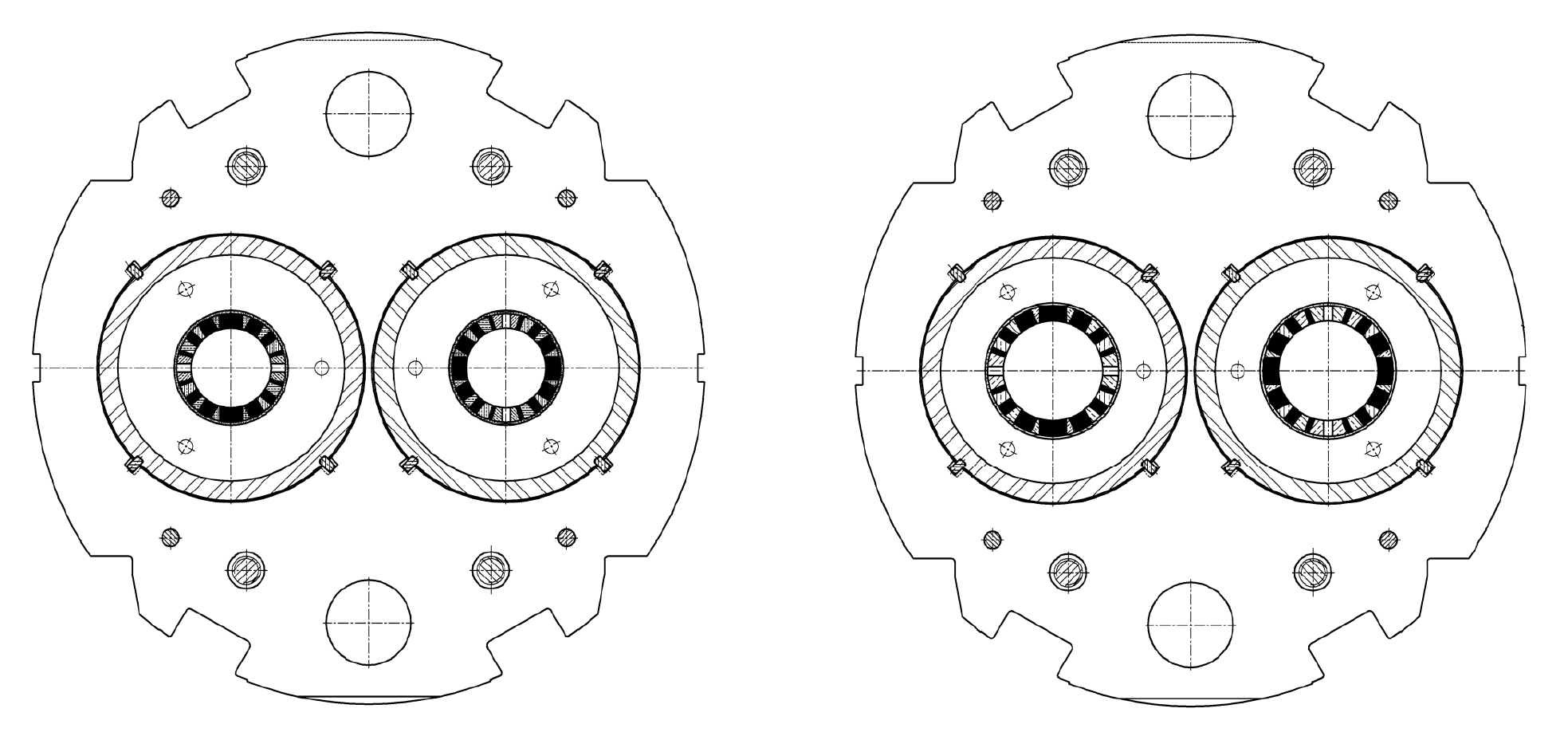}
}
 \caption{\label{fig:cross-section}%
 Cross-section of the MCBC (left) and the MCBY (right).}%
\end{figure}

The MCBC and MCBY magnets are 1.1-m long, double-aperture Nb--Ti dipole magnets able to reach a field level of 2.3-3.1~T. The magnets consist of two superconducting dipole modules characterised by a bore diameter of 56 mm (MCBC) and 70~mm (MCBY), which are mounted in a common support (Fig.~\ref{fig:cross-section}). For each magnet, looking from the connection side, the aperture on the left provides a horizontal field, while independently the aperture on the right provides a vertical field. Both magnet types use the same superconducting wire with a rectangular cross-section  (0.38~mm $\times$ 0.73~mm). The dipole coils are wound from flat ribbons of either 14 (MCBC) or 15 (MCBY) wires. The parameters of the MCBC and MCBY orbit correctors are given in Table~\ref{tab1}.

\begin{table}[htb!]
\begin{center}
\begin{minipage}{300pt}
\caption{Main parameters of the MCBC and MCBY orbit correctors~\cite{Bruning:2004}}\label{tab1}%
\begin{tabular}{@{}llll@{}}
\toprule
Parameter & Unit & MCBC & MCBY\\
\midrule
Coil inner diameter    & mm   & 56  & 70  \\
Magnetic length    & m   & 0.904  & 0.899  \\
Operating temperature     & K   & 1.9/4.5  & 1.9/4.5  \\
Design field at 1.9/4.5 K       & T      & 3.11/2.33 & 3.00/2.50  \\
Design current at 1.9/4.5 K     & A      & 100/74    & 88/72  \\
Max. operating current at 4.5K  &       &           &           \\
~~~during 2018 vdM scans           & A     & 80        & 77        \\
Superconductor type   & --   & Nb--Ti in Cu matrix & Nb--Ti in Cu matrix \\
Wire dimension   & --   & 0.38~mm $\times$ 0.73~mm & 0.38~mm $\times$ 0.73~mm \\
Ribbon construction  & --   & 14 wires (glued) & 15 wires (glued)\\
\botrule
\end{tabular}
\end{minipage}
\end{center}
\end{table}

Magnetic measurements of spare magnets of type MCBC and MCBY were carried out at the cryogenic test station of SM18 at CERN. A rotating-coil system~\cite{Arpaia:2012_1, Arpaia:2012_2}, that provides a measurement of the main field (and therefore of the transfer function) with a typical precision of 0.01$\%$ and an accuracy of 0.1$\%$, was used to characterize the integral transfer function~\cite{DAVIES}. Each aperture was equipped with a rotating coil shaft composed of 5 segments, each 0.223~m long. The measurements were conducted at a temperature of 4.5~K, since this is the operating temperature for both orbit correctors in the LHC.

\subsection{\label{sec3:2}Reference cycles}

\begin{figure}[htb!]
\centering
{\includegraphics[scale=0.75]{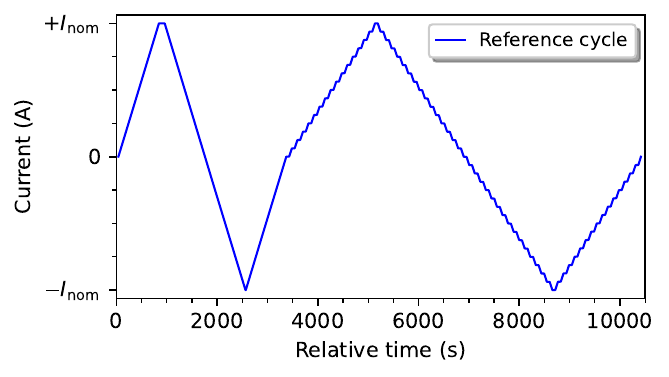}
}
 \caption{\label{fig:ref}%
  Powering diagram of a reference cycle.}%
\end{figure}

A first set of measurements consisted of full-range current cycles, denoted \textit{reference cycles}, which are used for a first characterization of the magnetic field of superconducting magnets. The aperture under test was powered up to the maximum positive and then to the maximum negative nominal current $I_{\mathrm{nom}}$, which is $80$~A for MCBC and $77$~A for MCBY. This pre-cycle was followed by a stair-step profile (Fig.~\ref{fig:ref}). The current in the adjacent aperture was set to zero.

\subsection{\label{sec3:3}VdM cycles}

A second set of measurements included \textit{vdM cycles}, emulating the powering conditions during luminosity-calibration scans by means of vdM and LSC methods~\cite{Chmielinska:2022}. For brevity, we will only describe the vdM cycles for MCBC corresponding to the powering schemes performed during LHC fill 6016, shown in Fig~\ref{fig:6016}. The first cycle emulated the actual powering of the circuit with the highest current amplitude, while the second emulated the powering of the circuit with the lowest current amplitude. To reduce the measurement duration, we omitted the long plateaus of constant current.

\begin{figure}[htb!]
\centering
{\includegraphics[width=1\linewidth]{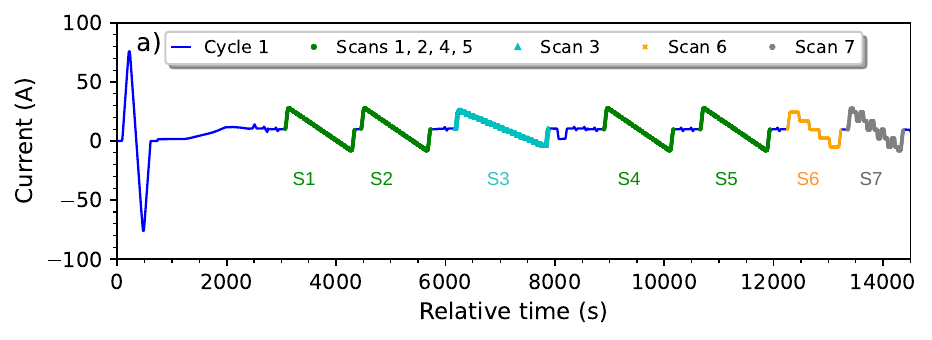}}
\\
{\hspace*{0.25cm}
\includegraphics[width=0.31\linewidth]{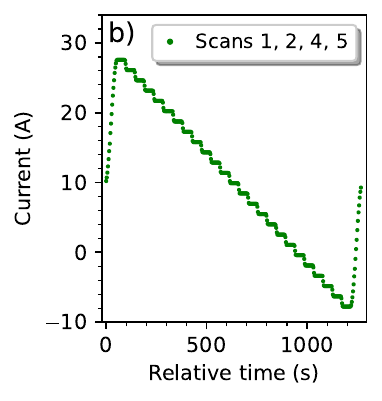}}%
\includegraphics[width=0.31\linewidth]{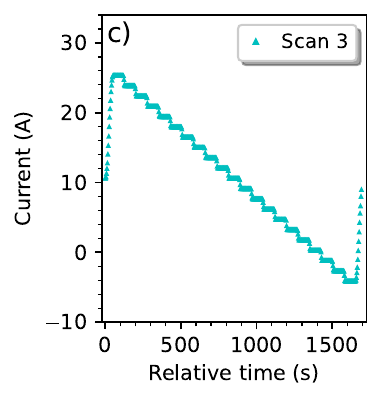}
{\includegraphics[width=0.31\linewidth]{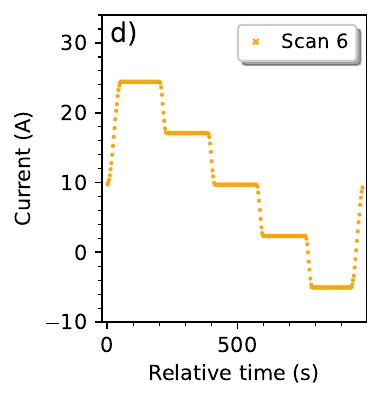}}%
 \quad 
 \caption{\label{fig:SP1}%
 (a) Powering diagram of the first vdM cycle. Details of scans with a monotonic decrease of the current, including: (b) vdM scans, (c) scan 3, (d) scan 6.}%
\end{figure}

The powering diagram of the first vdM cycle is displayed in Fig.~\ref{fig:SP1}a. The cycle consists of an initial pre-cycle followed by vdM scans (1, 2, 4 and 5). The powering levels for vdM scans are identical, with the current decreasing from 27.5~A to $-7.8$~A in 25 steps (Fig.~\ref{fig:SP1}b). Scans 3 and 6 have a different powering range and fewer intermediate levels (Figs.~\ref{fig:SP1}c and~\ref{fig:SP1}d). 

\begin{figure}[H]
\centering
{\includegraphics[width=1\linewidth]{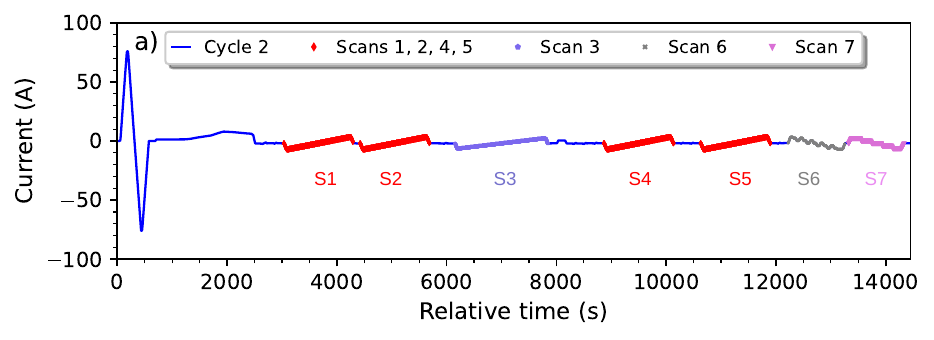}} \\
{\hspace*{0.25cm}
\includegraphics[width=0.295\linewidth]{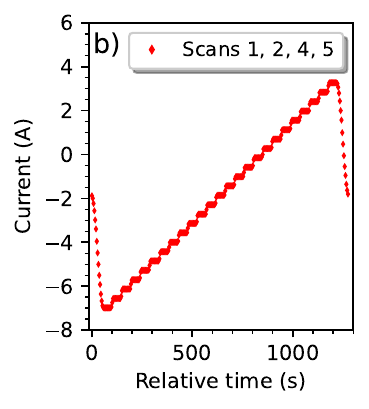}}%
 \quad 
\includegraphics[width=0.295\linewidth]{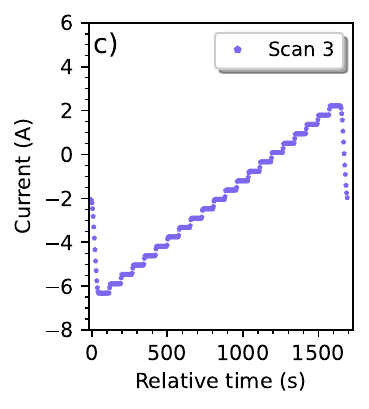}
% {\includegraphics[width=0.48\linewidth]{output_39_4}}%
% \quad
\includegraphics[width=0.31\linewidth]{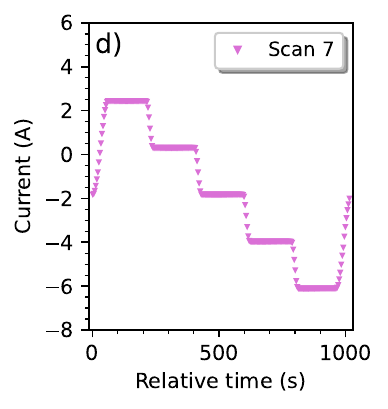}\\
 \caption{\label{fig:SP2}%
(a) Powering diagram of the second vdM cycle. Details of scans with a monotonic increase or decrease of the current, including: (b) vdM scans, (c) scan 3, (d) scan 7.}%
\end{figure}

The powering diagram of the second vdM cycle is shown in Fig.~\ref{fig:SP2}a. The cycle consists of an initial pre-cycle followed by vdM scans (1, 2, 4 and 5), detailed in Fig.~\ref{fig:SP2}b. For vdM scans, the current increases from -7.0~A to 3.3~A in 25 steps. Scan 3 has a different powering range and fewer intermediate steps (Fig.~\ref{fig:SP2}c). The same applies to the scheme of scan 7, which contains an opposite current-ramp (Fig.~\ref{fig:SP2}d).

Both vdM powering schemes contain also an LSC scan, which includes multiple changes in ramp direction, \textit{e.g.} scan 7 in  Fig.~\ref{fig:SP1}a and scan 6 in Fig.~\ref{fig:SP2}a. These scans will be discussed in Sec.~\ref{sec4:2:2}.

\section{\label{sec4}Analysis of magnetic measurements}

The analysis of the magnetic measurements requires advanced post-processing of the raw data, \textit{i.e.} of the induced voltage signals in the magnetometers. Unless stated otherwise, we performed this analysis after filtering the data acquired at the current plateaus, since the measurement precision at constant current level is better than 0.01$\%$. At the nominal current, the integral field resulted in 2.2677~Tm for MCBC and 2.4130~Tm for MCBY. For both magnets, we evaluated the \textit{geometric term}, which describes the linear dependence of the field on the operating current~\cite{Sammut:2009}. For this purpose, a linear fit was computed for selected data points of a reference cycle and within a linear range, below saturation, of $\pm$29~A (MCBC) and $\pm$25~A (MCBY). The coefficients of the linear fit were 0.02862~Tm/A (MCBC) and 0.03260~Tm/A (MCBY). 
 The nonlinearity from persistent currents has been retrieved as the residual of the full set of data-points with respect to the geometric term of a specific magnet. 

\subsection{\label{sec4:1}The reference magnetization loop}

\begin{figure}[htb!]
\centering
\includegraphics[scale=0.70]{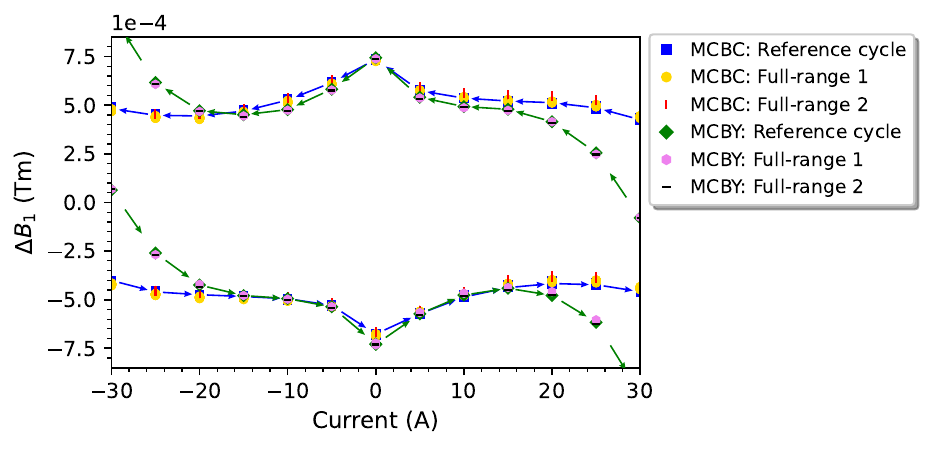}% Here is how to import EPS art
\caption{\label{fig:ComparisonStd} $\Delta B_{1}$ evaluated for MCBC and MCBY magnets in full-range cycles.}
\end{figure}

The residuals after subtracting the linear term for full-range cycles are denoted $\Delta B_{1}$ and shown in Fig.~\ref{fig:ComparisonStd}. At low field levels, the hysteresis is visible, which results from cycling the magnet up to the maximum positive and the maximum negative
current. This excitation cycle is referred to as a \textit{major magnetization loop}. As a consequence, the magnetic-field value depends on the ramp direction, which is indicated by the arrows in Fig.~\ref{fig:ComparisonStd}. For both magnets, the nonlinearity at low field levels is similar, as they use the same wire type. The hysteresis half-width is $\pm$0.74~mTm (0.031$\%$ relative to the nominal field) for MCBC, and $\pm$0.75~mTm ($\pm$0.033$\%$ relative to the nominal field) for MCBY. In both cases, the amplitude of the nonlinearity lies well below the $1\%$ required by standard LHC operation. In some applications, however, a more accurate prediction of the magnetic field is required. We have therefore updated the FiDeL model of both magnets by fitting new measurement data~\cite{Chmielinska:2022}.

At high field levels, the saturation of the iron yoke contributes to the nonlinearity of the magnet. This effect is particularly visible for the MCBY magnet outside the linear range of $\pm$25~A, in Fig.~\ref{fig:ComparisonStd}. The saturation of MCBY was found to be $\pm$97.2 mTm at nominal current, $\sim$4.5 times larger than for MCBC, due to a different coil cross section. However, since these current values are not reached during the vdM scans, the saturation component does not play a significant role for the physics discussed in this paper.

To evaluate the effect of cross-talk, we analyzed two additional full-range cycles, where the adjacent aperture was powered with either the positive or the negative maximum possible current. For MCBC, the largest difference between the residuals of the reference cycle (Fig.~\ref{fig:ComparisonStd}, blue squares) and the two additional full-range cycles (Fig.~\ref{fig:ComparisonStd}, yellow dots and vertical red dashes) was $\pm0.01$~mTm ($\pm0.001\%$ in relative terms), implying that the effect of cross-talk is negligible. The same holds for the MCBY magnet.

\subsection{\label{sec4:2}VdM magnetization loops}

Nonlinear effects from persistent currents are particularly important from the perspective of accelerator operation with complex powering schemes, such as during vdM luminosity calibration scans. So-called \textit{vdM magnetization loops} are expected to be observed for cycles where the applied magnetic field is reversing, typically at field levels significantly below the nominal value. Of particular interest in such cases is the reproducibility of the magnetic field, both of the linear term and of the residual nonlinearities. In this Section, we present the results of magnetic measurements for the two vdM cycles described in Sect.~\ref{sec3}. In particular, we discuss the impact of the powering profile on the nonlinearity, focusing on cycles with either a monotonic current ramp (\textit{e.g.} vdM scans) or a reversing current ramp (\textit{e.g.} LSC scans).  

\subsubsection{\label{sec4:2:1}Cycles with monotonic ramps}

\begin{figure}[htb!]
\centering
\includegraphics[scale=0.70]{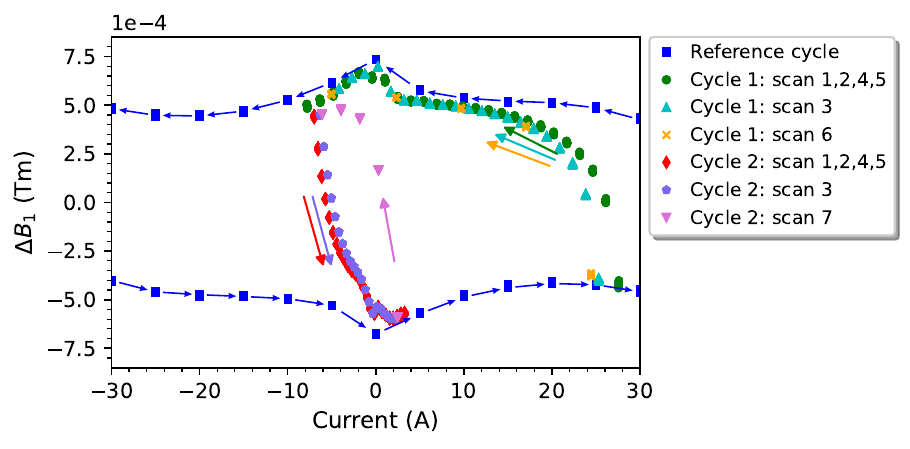}
\caption{\label{fig:Comparison} $\Delta B_{1}$ evaluated for the MCBC reference cycle and selected scans with monotonic current ramps of vdM cycle 1 and vdM cycle 2. The arrows indicate the current direction during the cycles.}
\end{figure}

At first, let us consider only scans with a monotonic increase or decrease of the current, shown in Figs.~\ref{fig:SP1}b-d and~\ref{fig:SP2}b-d.
We remark that the same colour notation and markers are used in the analysis of magnetic measurements.  
For the aforementioned scans, we computed residuals from the linear term and compared them in Fig.~\ref{fig:Comparison} to the residuals computed for the reference cycle of the MCBC magnet. A nonlinearity is apparent, that originates from the transitions between the branches of the major magnetization loop.

For scans 1 to 6 of the first cycle, all of which have a decreasing current amplitude (indicated by the corresponding arrows), a full transition from the lower to the upper magnetization branch occurs. For scans 3 and 6, the transition starts at a lower current, due to a different powering profile.

For scans 1 to 5 of the second cycle, all of which have an increasing current amplitude, there is a transition from the upper to the lower magnetization branch. For scan 3, the transition starts at a lower current compared to standard vdM scans, due to a different powering profile. For scan 7, the direction of the ramp is opposite to that in the other cycles and therefore the transition occurs from the lower to the upper branch.

For both vdM cycles, the peak-to-peak amplitude of the nonlinearity remains within the full width of the major magnetization loop. It is noteworthy that the data points corresponding to the vdM scans overlap, which demonstrates the repeatability of the results. Furthermore, the results remain consistent even when different scans take place in-between.

\subsubsection{\label{sec4:2:2}Cycles with inversions of the ramp direction}

\begin{figure}[htb!]
\centering
\includegraphics[scale=0.66]{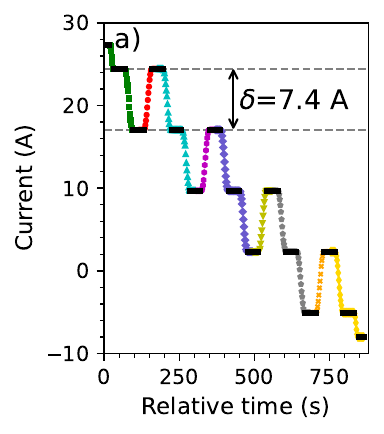}
\includegraphics[scale=0.68]{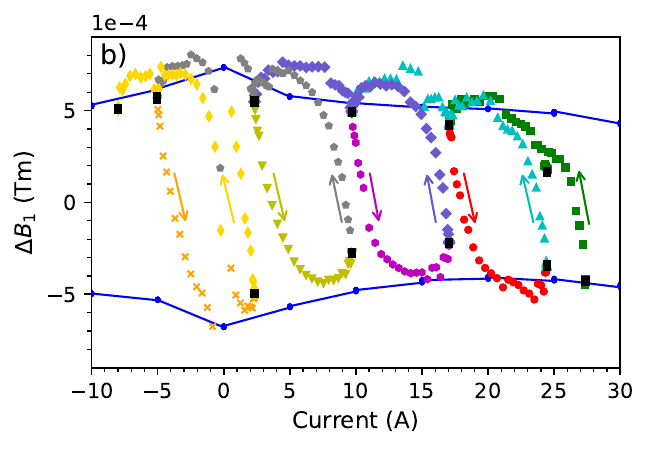} 
\caption{\label{fig:SP1scan7} Measurement results for scan 7 of vdM cycle 1: (a) current profile with the plateaus indicated by black markers and variations in the ramp direction indicated by different markers, (b) $\Delta B_{1}$ evaluated for the corresponding time intervals together with the major magnetization loop for the MCBC reference cycle indicated by the solid blue curve.}
\end{figure}

Let us now consider cycles that include inversions in the current-ramp direction. Figure~\ref{fig:SP1scan7}a depicts in full detail the powering profile of scan 7 of the first vdM cycle (see Fig.~\ref{fig:SP1}a). The corresponding residuals evaluated for all data points, including the ramps, are displayed in Fig.~\ref{fig:SP1scan7}b together with the major magnetization loop. The current is changed in steps of $\delta=7.4$~A; two consecutive steps in the down-ramp direction are followed by one step in the up-ramp direction. At each inversion of the ramp direction, a transition between magnetization branches occurs. In particular, an up-ramp immediately followed by a down-ramp at the same current level results in creating a minor magnetization loop. As can be seen in Fig.~\ref{fig:SP1scan7}b, a change of $\delta$ is typically sufficient to reach the lower branch of the major magnetization loop.

\begin{figure}[htb!]
\centering
\includegraphics[scale=0.65]{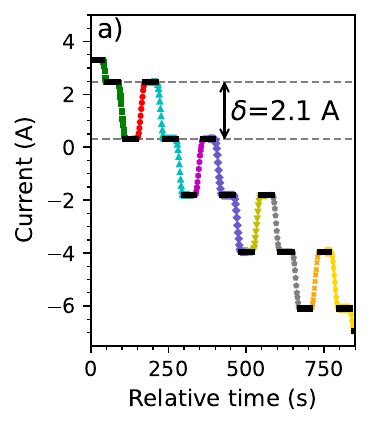}
\includegraphics[scale=0.65]{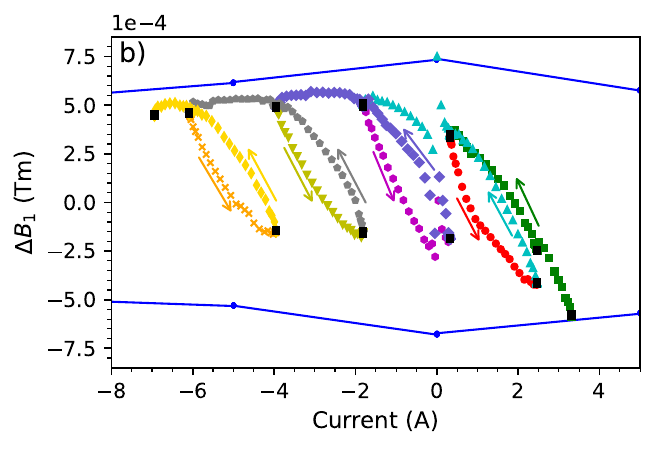} 
\caption{\label{fig:SP2scan6}  Measurement results for scan 6 of vdM cycle 2: (a) current profile with changes in the ramp direction indicated by different markers, (b) $\Delta B_{1}$ evaluated for the corresponding time intervals together with the major magnetization loop for the MCBC reference cycle indicated by the solid blue curve.}
\end{figure}

Figure~\ref{fig:SP2scan6}a shows the powering profile of scan 6 of the second vdM test cycle (see Fig.~\ref{fig:SP2}a). The residuals are analyzed separately and shown in Fig.~\ref{fig:SP2scan6}b. During this scan, the current is increased by $\delta=2.1$~A and decreased by $2\delta$, as shown in Fig.~\ref{fig:SP2scan6}a. In this case, after an inversion of the ramp direction, even a change of current by $2\delta$ is not sufficient for a full transition towards the major magnetization loop, as shown in Fig.~\ref{fig:SP2scan6}b, because the new shielding current layer has not fully penetrated the filament.

\subsection{\label{sec4:3}Repeatability of the linear coefficient}

A recurring, and sometimes dominant, source of systematic uncertainty in the absolute luminosity scale is the reproducibility of vdM calibrations: consecutive scans in the same LHC fill, recorded under supposedly identical conditions, can yield calibrations results that differ by 0.5\% or more~\cite{Atlas}. Several mechanisms have been found to contribute; among them, the non-reproducibility of beam orbits during the scans is a frequent culprit. This observation suggested to quantify the reproducibility of the field produced by the orbit correctors involved in the luminosity scans. As a guideline, the actual beam displacements at the IP, and therefore the field integrals of each of the relevant orbit correctors, should be reproducible to significantly better than 0.1\% of the maximum dialed-in beam displacement.

We analysed the linear relationship between the magnetic field and the current, separately for each of the scans performed during the vdM cycles. Let us denote the \textit{linear coefficient} of the $i^{\mathrm{th}}$ scan  by $\gamma_{i}$, and the \textit{average linear coefficient} ({\textit i.e.} the mean value of the $\gamma_{i}$'s) for all the scans of the $k^{\mathrm{th}}$ cycle by~$\overline\gamma_{k}$.

The average linear coefficients for the two vdM cycles were found to be  $\overline\gamma_{1}=0.02860$~Tm/A and $\overline\gamma_{2} = 0.02853$~Tm/A, a difference of 0.25\%. This is unlikely to affect the reproducibility of the beam displacements, since these two cycles mirror the powering history of two orbit correctors that play complementary roles within the closed-orbit bumps used for scanning. Such a difference in response will instead manifest itself by a small non-closure of these closed-orbit bumps, since two magnets that are supposed to be identical will exhibit slightly different transfer functions. Such ``bump leakage'' is of no concern as long as its magnitude is reproducible from scan to scan.

What matters instead is the repeatability of the linear coefficient under identical powering conditions. This can be quantified as the difference between this linear coefficient in a given scan and its mean value over all scans of the $k^{\mathrm{th}}$ cycle, normalized to that same mean value:
\begin{equation}\label{eq:repeatability}
    \xi_{k} = \frac{\gamma_{i} - \overline\gamma_{k} }{\overline\gamma_{k}}.
\end{equation}

\begin{figure}[htb!]
\centering
\includegraphics[width=0.38\linewidth]{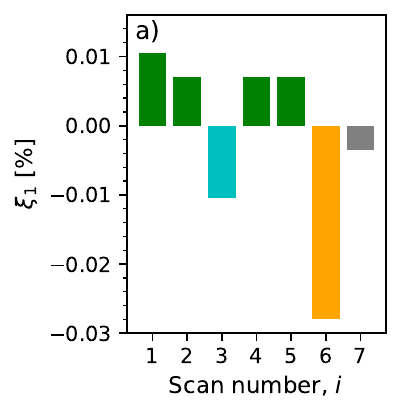}
\includegraphics[width=0.38\linewidth]{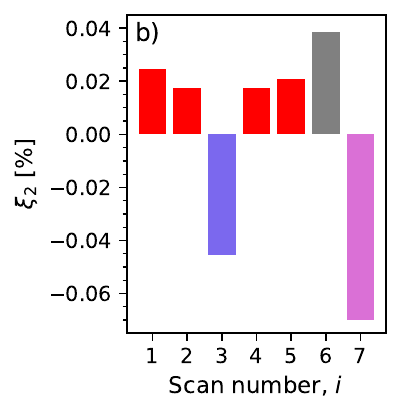} 
\caption{\label{fig:reprod} Repeatability of the linear coefficient during (a) vdM cycle~1 and (b) vdM cycle~2. The colours for the individual scans correspond to those used in Figs.~\ref{fig:SP1} and \ref{fig:SP2}.}
\end{figure}

\noindent The results are displayed in Fig.~\ref{fig:reprod} for the two vdM cycles separately. 

In the first cycle (Fig.~\ref{fig:reprod}a), the maximum difference between linear coefficients during vdM scans only (green towers) is 0.003$\%$ when normalized to the mean value, well within the above requirements. Considering all the scans, the maximum difference occurs between scans 1 and 6, and amounts to 0.038$\%$ in relative terms. This suggests that when different scan types take place within a luminosity-calibration session, the scan-to-scan reproducibility could be improved by inserting, before each scan, a ``mini-standardization cycle'', akin to the standard cycle described in Sect.~\ref{sec3:2} but with an amplitude compatible with the maximum tolerable closed-orbit excursion.

In the second cycle (Fig.~\ref{fig:reprod}b), the maximum difference in linear coefficients across vdM scans is 0.007$\%$ in relative terms, again well within requirements. Considering all the scans, the maximum difference between linear coefficients (scans 6 and 7) is 0.109$\%$ in relative~terms.

\subsection{\label{sec4:4}Implications for accelerator operation}

The results presented above demonstrate that the magnetization-branch transition is the main source of nonlinearity at low field levels in vdM powering conditions. For cycles with monotonic ramps, a single transition occurs between the reference magnetization branches. In turn, the inversion of the ramp direction causes the shift towards the opposite branch, which results in the creation of a minor magnetization loop. The observed transitions are not instantaneous. For a complete transition, a current change of about 8-10~A is required. This effect requires special attention when an accurate prediction of the magnetic field is needed. As already pointed out, the FiDeL system predicts magnetization effects only for full-range cycles, involving modelling of the major magnetization loop. For the vdM cycles, predicting the magnetization-branch transitions is particularly challenging and will be carried out in the next section.

\section{\label{sec5}Numerical model of the magnetic hysteresis}

A description of the individual branch transitions is necessary to achieve greater precision in the magnetic-field prediction for deflecting the beam in the transverse plane. However, it is not possible to characterise all machine cycles experimentally. In the present work, we have further investigated the relationship between the current change required to achieve a full transition and the initial current at which inversion starts, in order to develop a numerical model of the magnetic hysteresis.

\subsection{\label{sec5:1}Parametric fitting}
\begin{figure}[htb!]
\centering
\includegraphics[scale=0.67]{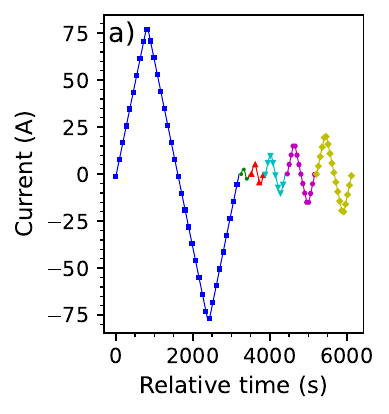}
\includegraphics[scale=0.65]{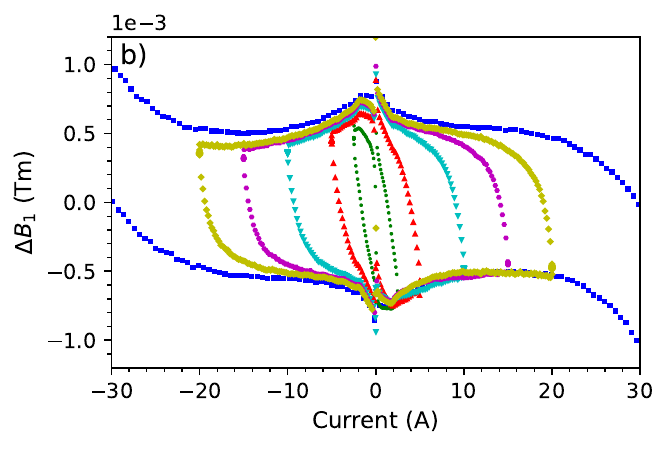} 
\caption{\label{fig:Fig9} Measurement results used to model magnetization-branch transitions: (a) powering diagram of the MCBY magnet (individual cycles are distinguished by different markers and colours), (b) $\Delta B_{1}$ evaluated for the corresponding cycles. }
\end{figure}

For this purpose, we designed an additional experiment, in which the MCBY magnet was powered with a pre-cycle followed by a sequence of up-ramp and down-ramp cycles, each with a different amplitude, as shown in Fig.~\ref{fig:Fig9}a. For those cycles, the residuals from the linear term are displayed in Fig.~\ref{fig:Fig9}b.

Experimental results show that exponential transitions occur for cycles in which the sign of the current ramp is reversed at a certain current level, \textit{i.e.} above $\pm5$~A. At lower currents, the transitions are not exponential or do not reach the reference magnetization loop. We modelled exponential transitions by fitting the measurement data with the following function:

\begin{equation}
\Delta {B_{1}} (I, \mathrm{sgn}(\frac{dI}{dt})) = ae^{-\mathrm{sgn} (\frac{dI}{dt}) b I} + c,
\label{eq:parametricFit}
\end{equation}

\noindent where $\mathrm{sgn} (\frac{dI}{dt})$ indicates the sign of the current ramp (positive for up-ramp and negative for down-ramp), 
$a$ (expressed in Tm) and $b$ (expressed in A$^{-1}$) are the fitting coefficients, $c$ (expressed in Tm) is the amplitude of the plateau of the major magnetization loop at low currents. 

\begin{figure}[htb!]
\centering 
\includegraphics[scale=0.65]{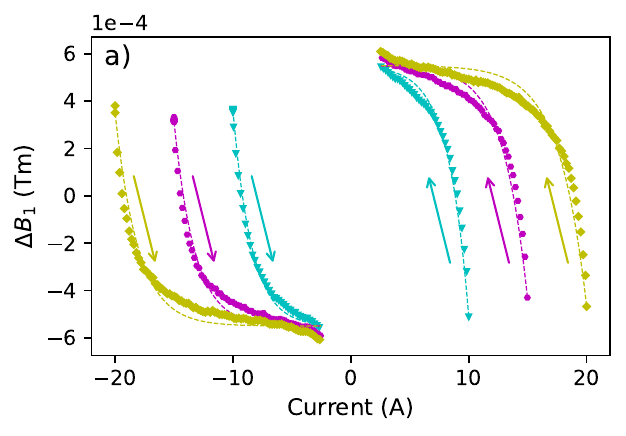}
\includegraphics[scale=0.65]{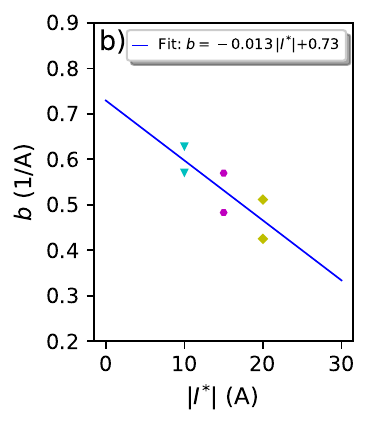} 
\caption{\label{fig:Fig10} Parametric fitting of the magnetization-branch transitions: (a) comparison of the experimental data and the fitting model (the arrows indicate the ramp direction), (b) the current dependency of the fitting parameter~$b$ (data points below the linear fit correspond to the down-ramp, while the data points above the linear fit correspond to the up-ramp). }
\end{figure}

A comparison of the measurement results with the fits for individual cycles is presented in Fig.~\ref{fig:Fig10}a. For parameter $a$, we have not observed any obvious pattern. However, the $b$ parameter depends linearly on the current level at which the ramp reversal occurs ($I^{*}$), as displayed in Fig.~\ref{fig:Fig10}b. Therefore, we can describe $b$ with a linear function: $b(I^{*})=-0.013\, \lvert I^{*} \rvert +0.73$. An analogous dependence is present when analyzing the measurement results of the MCBC magnet. In this case, the current dependence of the parameter $b$ is: $b(I^{*})=-0.011\,\lvert I^{*} \rvert+0.55$.

The method described thus far constitutes an empirical approach to the modelling of the branching in magnetic hysteresis. From a physical point of view, however, the magnetisation is not a function of the transport current, but depends on the applied magnetic field. Hence, to validate our model, we analyzed the relation:

\begin{equation}
\Delta {B_{1}} (B_{1}, \mathrm{sgn}(\frac{dI}{dt})) = ae^{-\mathrm{sgn} (\frac{dI}{dt}) h B_{1}} + c,
\label{eq:parametricFitField}
\end{equation}
where $h$ is the fitting coefficient (expressed in Tm$^{-1}$). It turns out that the parameter $h$ can be modelled as $h(B_{1}^{*})=-12.82\, \lvert B_{1}^{*} \rvert +19.23$ for the MCBC magnet and $h(B_{1}^{*})=-12.63\, \lvert B_{1}^{*} \rvert +22.63$ for the MCBY magnet, where $B_{1}^{*}$ indicates the magnetic-field level at which reversion occurs. The fitting parameters are similar, suggesting that the response of the superconductor to the external magnetic field is comparable for both magnets. This is the expected behaviour, since the MCBC and MCBY magnets are made of the same wire type.

\subsection{The numerical model}

The FiDeL model predicts the magnetic-field value on the major magnetization loop based on the ramp direction and the current~\cite{Sammut:2006}. Let us denote the magnetization contribution as $\Delta B_{1}^{\mathrm{FiDeL}}(I, \mathrm{sgn} (\frac{dI}{dt}))$. Here, we aim to extend the FiDeL model by predicting the  magnetization-branch transitions for arbitrary cycles. A single transition occurring from $I^{*}$ to 0~A  can be modelled using the following formula:

\begin{equation}
\Delta {B_{1}} (I, \mathrm{sgn} (\frac{dI}{dt})) = ae^{-\mathrm{sgn} (\frac{dI}{dt}) b(I^{*}) I} + c,
\label{eq:transition}
\end{equation}

\noindent where $b(I^{*})$ is the fit function of a specific magnet described in paragraph~\ref{sec5:1} and $c$ is the amplitude of the plateau of the major magnetization loop predicted by the FiDeL model. To determine the last parameter $a$, we use the magnetic-field value predicted from the FiDeL model evaluated at $I^{*}$ ($\Delta B_{1}^{\mathrm{FiDeL}}(I^{*}, \mathrm{sgn} (\frac{dI}{dt})$) and derive $a$ directly:

\begin{equation}
a = \frac {\Delta B_{1}^{\mathrm{FiDeL}}(I^{*}, \mathrm{sgn} (\frac{dI}{dt})) - c}{e^{-\mathrm{sgn} (\frac{dI}{dt})  b(I^{*}) I^{*}} }.
\end{equation}

\begin{figure}[htb!]
\centering
\includegraphics[scale=0.75]{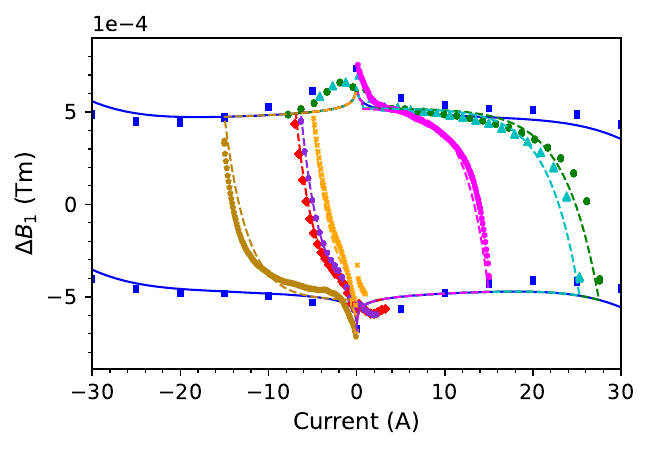}% Here is how to import EPS art
\caption{\label{fig:FiDeLExtended} Comparison of the measured magnetization-branch transitions and the proposed extended FiDeL model for the MCBC magnet. The measurement results are indicated by markers, the numerical-model results by solid and dashed curves.}
\end{figure}

We applied the new model to a separate measurement dataset for the MCBC magnet. The comparison between the measured magnetization-branch transitions and the predictions of the model is shown in Fig.~\ref{fig:FiDeLExtended}. The maximum difference between the measurement results and the model is $\pm0.21$~mTm (0.009$\%$ relative to the nominal field).

\section{\label{sec6}Impact of hysteresis on beam-orbit distortions}
The change of a magnetic dipole field $\Delta B_{1}$ yields a dipole kick $\Delta k$ \cite{Chao}:
\begin{equation}
\Delta k = \Delta {B_{1}}\,/R\label{eq:rigity},
\end{equation}
where $R$ is the magnetic rigidity of the beam.

The change of the beam closed-orbit position $\Delta u(s)$ at a longitudinal position $s$ in plane $u = x,y$ due to a dipole kick $\Delta k_u$ in this plane at position $s_k$ is given by \cite{Chao}:
\begin{equation}
\Delta u(s)=\frac{\sqrt{\beta_{u}(s) \beta_{u}(s_k)}\,\cos(\lvert \mu_{u}(s)-\mu_{u}(s_k) \rvert - \pi Q_u)}{2 \sin(\pi Q_{u})}\Delta {k_u},\label{eq:orbit}
\end{equation}
where $\beta_{u}$ is the optical beta function, $\mu_{u}$ is the phase advance, and $Q_u$ is the machine tune.
Since $\Delta u$ is linear in $\Delta k_u$, the combined effect of multiple dipole kicks is the sum of the individual kick responses.

\begin{figure}[htb!]
\centering
\includegraphics[width=1\linewidth]{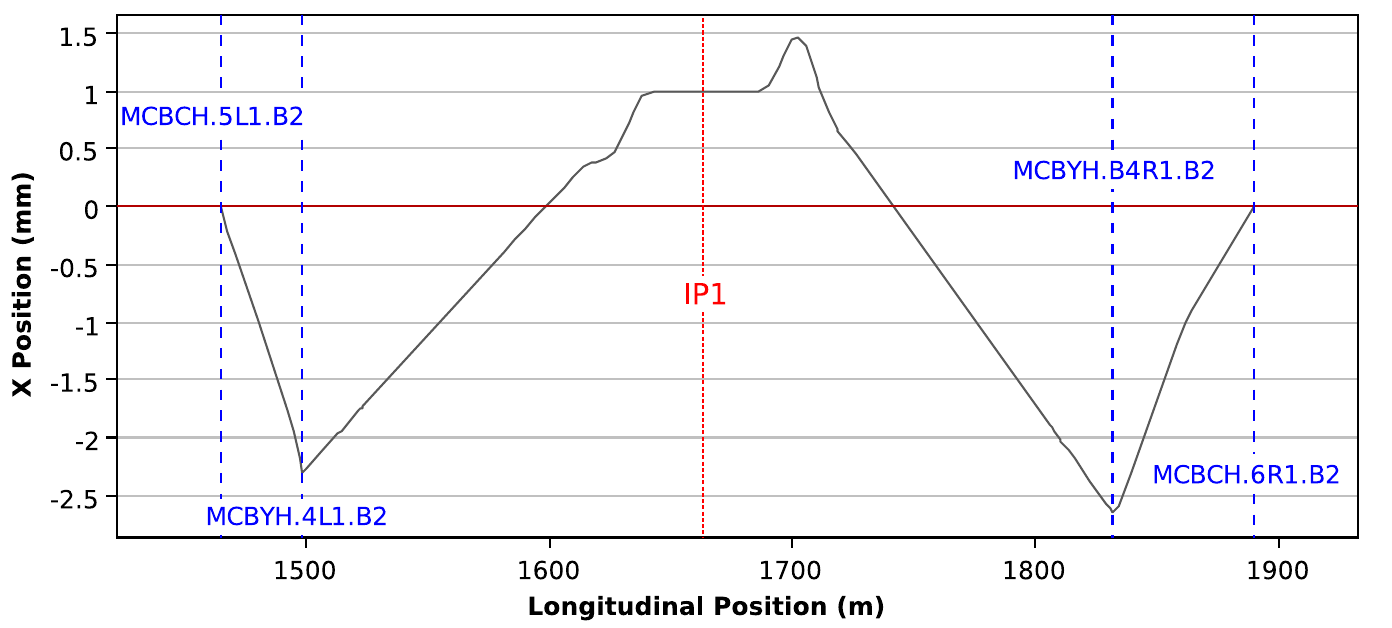}
\caption{\label{fig:bumpIP1B2X} $1\ut{mm}$ closed-orbit bump of Beam 2 at LHC IP1 in the horizontal plane for LHC injection optics. The four magnets involved (2 MCBC and 2 MCBY) are marked in blue; IP1 is marked in red.}
\end{figure}

Using four dipole orbit-corrector magnets with relative phase-advances optimized for this purpose, a closed-orbit bump that controls both the beam position and the beam angle at a specific location around the machine can be built. Such closed-orbit bumps are used to control the position of each beam, in each plane at the four LHC IPs using the MCBC and MCBY magnets. Due to the linearity of Eq.~\ref{eq:orbit}, they can be scaled by applying a scaling factor to all kicks. These bumps are then used to displace the beams during luminosity-calibration scans, as outlined in Sect.~\ref{sec2:3}. An example is given in Fig.~\ref{fig:bumpIP1B2X}.

The total impact of the magnetic hysteresis from persistent currents on the beam position at the IP is derived by combining the standard FiDeL model with the numerical model of superconductor magnetization presented in Sect.~\ref{sec5} for the four magnets involved, calculating the closed-orbit change introduced by each magnet using Eq.~\ref{eq:orbit}, and summing the effects of the four magnets.

This approach can then be applied to a majority of LHC luminosity calibration scanning cycles to predict the additional beam position offsets introduced by magnetic hysteresis from persistent currents, based on the measured current history in the magnets involved.

For validation, this prediction is compared to beam displacements measured during a LSC scan, with the LHC in the injection configuration at a beam energy of $450\ut{GeV}$. The low beam-energy configuration is chosen for comparison both because the hysteresis effects are stronger at low energy due to the shallowness of the magnetic cycles during the scanning sequence (Sect.~\ref{sec4:2}), and because their impact on the beam trajectory is larger due to the lower magnetic rigidity (Eq.~\ref{eq:rigity}).

\begin{figure}[H]
\centering
\includegraphics[width=0.96\linewidth]{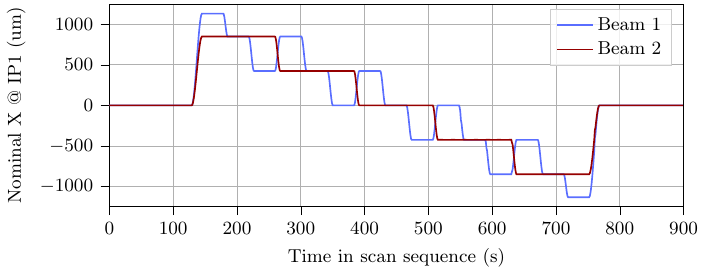}
\caption{\label{fig:LSCpos} Nominal beam displacements at LHC IP1 in the horizontal plane during a Length Scale Calibration session in 2018 (LHC fill 7300).}
\end{figure}

\begin{figure}[H]
\centering
\includegraphics[width=0.96\linewidth]{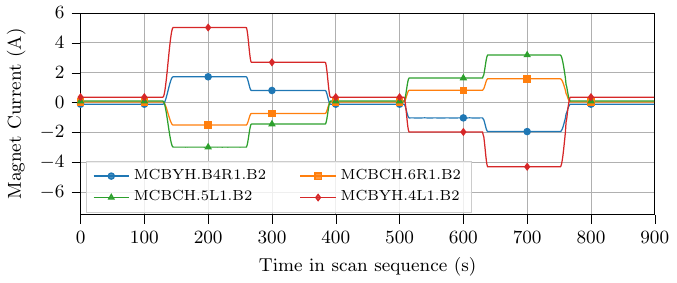}
\caption{\label{fig:LSCcurr} Currents in the four magnets that form the closed-orbit bump for Beam 2 in the horizontal plane at LHC IP1 during a Length Scale Calibration session in 2018 (LHC fill 7300).}
\end{figure}

\begin{figure}[H]
\centering
\includegraphics[width=0.96\linewidth]{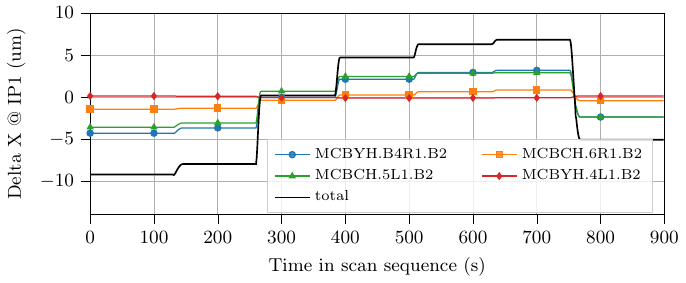}
\caption{\label{fig:LSCoff} Predicted contribution of each of the four magnets involved in the closed orbit bump to the hysteresis-introduced position offset of Beam 2 in the horizontal plane at LHC IP1 during a Length Scale Calibration scan session in 2018 (LHC fill 7300).}
\end{figure}

The nominal beam displacements dialed-in during a horizontal LSC scan are shown in Fig.~\ref{fig:LSCpos}. The corresponding currents in the magnets used to establish the closed-orbit bump for Beam 2 are displayed in Fig.~\ref{fig:LSCcurr}. It is worth noting that these currents form a very shallow magnetic cycle, the peak amplitude of which is less than 10\% of the nominal current in the MCBC and MCBY magnets (Sect.~\ref{sec3}). The numerical model predicts beam offsets introduced by magnetic hysteresis at the level of $1\ut{\%}$ of the nominal beam displacement ($\pm\,8\ut{\mu m}$ over a beam displacement of $\pm\,850\ut{\mu m}$); the contribution of the four involved magnets and the total effect are shown in Fig.~\ref{fig:LSCoff}. 
% Fill 7300 2018-10-14

\begin{figure}[htb!]
\centering
\includegraphics[width=0.95\linewidth]{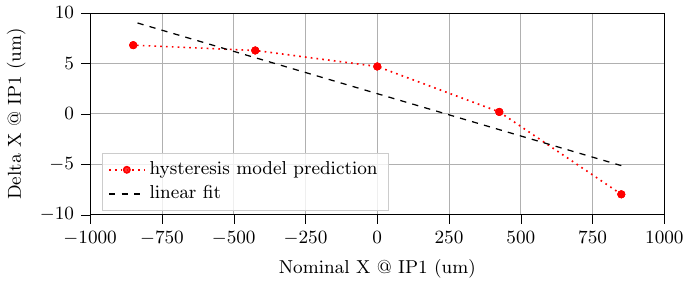}
\caption{\label{fig:LSCoffscan} Predicted total hysteresis-induced closed-orbit distortion of Beam 2 in the horizontal plane at LHC IP1, with respect to the nominal displacement of Beam 2 during a LSC scan session in 2018 (LHC fill 7300). The linear component (dashed line) is indistinguishable from a global length scale error and is subtracted  from the red points in the subsequent analysis.}
\end{figure}

During LSC scans, the beam positions at the IP are measured at each scan step in two independent ways: by interpolating the beam positions measured by the Beam Position Monitors (BPMs) at the final focusing magnets \cite{doros}; and by using the tracker of the experimental detector to measure the displacement of the luminous centroid~\cite{Atlas}. As described in Sect.~\ref{sec2:3}, the measured beam displacements are then compared to the requested nominal displacements, thereby establishing the absolute length scale of the four closed-orbit bumps. By construction of the LSC scans, any effect that causes a beam offset proportional to the nominal beam displacement is indistinguishable from an orbit-bump length-scale error and is absorbed in the LSC correction. As depicted by the linear fit in Fig.~\ref{fig:LSCoffscan}, the predicted beam position offsets from magnetic hysteresis contain such a linear component. In the following therefore, only the nonlinear component of the residuals of the measured beam positions and of the predicted hysteresis-induced offsets are compared.

In addition to the data collected during the 2018 LSC scans, a dedicated set of beam scans were performed in 2021 at IP1 of the LHC with the specific purpose to quantify the effect of magnetic nonlinearities and of hysteresis effects. For these \textit{nonlinearity} (NL) scans, noncolliding beams of longitudinally separated bunches were used to avoid beam-beam deflections \cite{Atlas}. The beams were scanned once parallel to each other (Beam 1 and Beam 2 displacements of the same sign), as in LSC scans, and once introducing a symmetric separation (Beam 1 and Beam 2 displacements of opposite sign), as in vdM scans. The beam positions during the scans were interpolated from the BPM measurements on either side of the IP. 

\begin{figure}[H]
\centering
\includegraphics{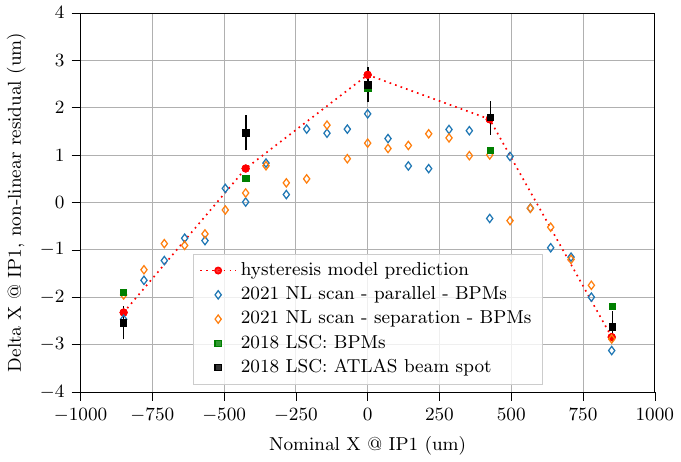}
\caption{\label{fig:IP1HistOrbit} Predicted and measured impact of magnetic hysteresis on the transverse displacement of Beam 2 at IP1 of the LHC, with respect to the nominal, dialed-in displacement during luminosity-calibration scans at 450 GeV. Only the nonlinear component of the residuals is shown; the linear component has been subtracted.}
\end{figure}

The two data sets (2018 LSC and 2021 NL scans) are described in detail in Ref.~\cite{Atlas}, and compared to the model prediction in Fig.~\ref{fig:IP1HistOrbit}. The predicted nonlinear part of the hysteresis-induced closed-orbit distortions are at the level of $0.3\ut{\%}$ of the nominal beam displacement ($\pm 2.5 \ut{\mu m}$ over a beam displacement of $\pm 850 \ut{\mu m}$), and well compatible with the measured displacements in both data sets. Magnetic nonlinearities of similar magnitude have been  observed in other luminosity-calibration scan sessions at the LHC; their impact on the absolute luminosity scale has so far been accounted for by an additional systematic uncertainty of up to 0.8\%~\cite{Atlas, CMS}.

\section{\label{sec7}Conclusions}
 
In this paper we analysed the nonlinear effects in superconducting accelerator corrector magnets due to superconductor magnetization at low currents. To characterize these phenomena, we designed a series of dedicated experiments, including full-range current cycles (reference cycles) and vdM cycles (cycles used during the vdM luminosity scans). For the latter ones, particular attention has been given to reproduce the relevant features of the powering sequences. During such scans, the superconducting orbit-corrector magnets MCBC and MCBY  are used in a parameter space that is far from that considered during their design. In particular, the magnets operate at low current (typically below $30\%$ of nominal value), with frequent variations of the current-ramp direction, which leads to magnetic hysteresis.

The experimental results show that nonlinear magnetization effects remain well below the accuracy level required for standard LHC operation ($< 1$\% relative to the main field), and that the reproducibility of the linear coefficient during identical vdM-like powering cycles is better than 0.01\%, comfortably within luminosity-scan requirements. The hysteresis half-width at zero current amounts to $\pm$0.74 mTm ($\pm$0.033$\%$ with respect to the nominal maximum field) for MCBC, and to $\pm$0.75~mTm ($\pm$0.031$\%$) for MCBY. In fact, as long as the excitation remains below saturation, the transitions between the branches of the major magnetisation loop are the main source of nonlinearity at low field levels. After an inversion of the current-ramp direction, the magnetization tends to stabilize on the major branch with a slow transition with an exponential profile. This phenomenon must be taken into account if more accurate predictions of the magnetic field are needed.

During luminosity-calibrations scans, the largest excitation of the corrector magnets is several times smaller than the nominal current associated with the main hysteresis loop. In order not to impact significantly the vdM-calibration uncertainty budget, the transfer function of these magnets must be known with higher precision (typically better than 0.1$\%$ with respect to the largest field excursion during the scans), and remain reproducible at the same level or better. Hence, not only are these magnets used in a peculiar operating range, but also the precision required of the FiDeL model is much more demanding compared to what is usually needed for corrector magnets. Moreover, the FiDeL system does not currently include the modelling of minor magnetization loops and of hysteresis-branch transitions. In this paper, we have presented a novel approach to fit the measurements of the aforementioned effect. This is crucial for predicting the beam displacements to better than 1$\%$ of the vdM- or LSC-scanning range.

Applying the numerical model of magnetic hysteresis proposed in this paper to data collected during luminosity-calibration scans in 2018 and 2021 has shown beam-positioning errors at the level of $0.3\ut{\%}$ ($\pm 2.5 \ut{\mu m}$ over a nominal beam displacement of $\pm 850 \ut{\mu m}$) due to hysteresis-branch transitions over the course of a scan. This is in good agreement with the beam-displacement data measured during these scans, and has so far been treated as a systematic uncertainty by the LHC experiments.

Furthermore, the analysis shows that it is not necessary to carry out multiple measurements of machine cycles to model the magnetic hysteresis. It is sufficient to experimentally study several up-ramp and down-ramp cycles, each with a different current amplitude, to investigate the linear relationship between the constant parameter of the exponential function describing the transition and the current level at which the transition occurs. Such information can be used to predict the magnetic field for most machine cycles, as illustrated in this paper.

\section*{Declarations}
\begin{itemize}
\item Availability of data and materials

The data presented in this article was generated at CERN. The raw data supporting the findings of these studies is stored in a database repository and is available on request from author L. Fiscarelli (lucio.fiscarelli@cern.ch).
\end{itemize}





\bibliography{Manuscript}% common bib file


\end{document}